\documentclass{raa}          
\usepackage{graphicx,times}             
\usepackage{natbib}
\usepackage{amssymb,amsmath}
\bibpunct{(}{)}{;}{a}{}{,}

\usepackage[pagebackref=true]{hyperref}

\begin{document}

  \title{The nature of elongated granulations and stretched dark lanes in a newly emerging flux region
}
   \volnopage{Vol.0 (20xx) No.0, 000--000}      
   \setcounter{page}{1}          

   \author{Jincheng Wang\inst{1,2}
   \and Xiaoli Yan\inst{1,2}
   }

   \institute{Yunnan Observatories, Chinese Academy of Sciences, Kunming 650011, People's Republic of China.; {\it wangjincheng@ynao.ac.cn}\\
        \and
             Yunnan Key Laboratory of Solar Physics and Space Science, Kunming 650011, People's Republic of China\\
\vs\no
   {\small Received 20xx month day; accepted 20xx month day}}

\abstract{In this study, we explore the elongated granulations and stretched dark lanes within the emerging anti-Hale active region NOAA 12720. Utilizing high-resolution observations from the New Vacuum Solar Telescope, we discern a prevalence of elongated granules and stretched dark lanes associated with the emergence of new magnetic flux positioned between two primary opposing magnetic polarities.
These elongated granulations and stretched dark lanes exhibit an alignment of strong transverse fields and a significant inclination angle. The endpoints of these features separate from each other, with their midpoints predominantly characterized by blue-shifted signals in the photosphere. This suggests a close association between elongated granules and stretched dark lanes with the newly emerging flux.
Additionally, we find that the stretched dark lanes display a more pronounced correlation with strong blue shifts and photospheric transverse magnetic fields compared to the elongated granulations. The transverse magnetic field within these stretched dark lanes reaches magnitudes of approximately 300 to 400 G, and the inclination angle demonstrates an ``arch-like'' pattern along the trajectory of the stretched dark lane.
Based on these observed characteristics, we infer the presence of an emerging flux tube with an ``arch-like'' shape situated along the stretched dark lane. Consequently, we conclude that the stretched dark lanes likely represent manifestations of the emerging flux tube, while the elongated granulations may correspond to the gaps between the emerging flux tubes.
}

   \authorrunning{J.-C. Wang et al. }            
   \titlerunning{Elongated granulations and stretched dark lanes}  

   \maketitle
   
\section{Introduction}           
\label{sect:intro}
Active regions are prominent and long-lasting features associated with the Sun's magnetic field, typically persisting for days, weeks, or even months. The prevailing understanding is that these active regions originate from the convergence of emerging magnetic flux \citep{van15, wan21}.
Initiated by convective flows and buoyant instabilities, the magnetic field beneath the solar surface can emerge in the form of $\Omega$-shaped loops into the solar atmosphere \citep{zwa85,cen07,iso08,che14}. As these new magnetic fields ascend from the subsurface into the solar atmosphere, they engage in interactions, either with ambient magnetic fields and plasma or among themselves (e.g., \citealp{xu22}). These interactions give rise to various observable signals in the atmosphere across diverse heights and temperature regimes.
These signals encompass elongated granulations \citep{cen17,kon20}, stretched dark lanes \citep{zwa85, wan20,wan20b}, UV bursts \citep{pet14, Tia18}, jets/H$\alpha$ surges \citep{liu04,wan18,wan19}, and Ellerman bombs \citep{ell17,rei16}.

Granules, characterized by irregular shapes and approximately 1-2 Mm in size, are pervasive on the quiet Sun's surface. Their lifetime ranges from 5 to 10 minutes, and their magnetic flux spans a considerable range, typically in the order of 10$^{15}$-10$^{18}$ Mx \citep{tit89,jin09}. 
In the granule interior, \cite{de02} found that small flux concentrations ($\sim$10$^{17}$ Mx) emerge and quickly disperse following the granule flows. In contrast to the quiet Sun's granules, those within an emerging flux region (EFR) exhibit distinct shapes. As the newly magnetic field emerges into the atmosphere from the subsurface, it interacts with these surface granules in intricate ways, inducing changes in morphology, altered plasma flows, and indications of heating \citep{cen17}.

However, based on a set of numerical radiative magnetohydrodynamics simulations, \cite{che07} concluded that magnetic flux tubes with less than 10$^{18}$ Mx of longitudinal flux are not sufficiently buoyant to rise coherently against the convective flows and produce no visible disturbances in granules. Elongated granulation is the one conspicuous phenomenon in the photosphere resulting from the interaction between emerging flux and photospheric plasma, which is generally considered as the indicator of emerging magnetic flux \citep{bra64}. Such aligned and elongated granulations coincide with the signature of a rising magnetic flux loop through the photosphere \citep{str96,sch10}. Using the Imaging Magnetograph Experiment (IMax) data, \cite{cen17} found that the elongated granulation in the direction of the main axis aligns with the direction of the emerging magnetic field. 
On the other hand, dark lanes have been observed in the photosphere through visible waveband observations, often correlated with a strong horizontal magnetic field \citep{lin11,wan20,sam19,wan20b, shen22}.  This type of dark lanes shows a straight and slender shape, which is quite different from the intergranular lane in the quiet Sun. Here, we call them stretched dark lanes. They are approximately aligned with the emerging magnetic flux, typically persist for about 10 minutes. Their darkness is attributed to the horizontal magnetic field suppressing turbulent heat exchange \citep{zwa85}. The appearance and disappearance of stretched dark lanes are associated with the apex of the rising loop through the photospheric layer \citep{sam19}.

Concentrating on magnetic emergence, we have explored the evolution and magnetic properties of active region NOAA 12702 in previous study \citep{wan21}. In this paper, we delve into the analysis of elongated granulations and stretched dark lanes associated with the magnetic emergence in this active region. Our aim is to comprehend the nature of these emerging observational features and shed light on the interaction between newly emerging flux and photospheric plasma. The sections of this paper are structured as follows: Section 2 outlines the observations and data reduction, Section 3 presents the results, and Section 4 provides the summary and discussions.

\section{Observations and data reduction}\label{sect:Obs}

The active region of interest gradually emerged near the center of the solar disk (e.g., N07, W16) between August 24 and 25, 2018. Throughout this period, the emerging active region was captured by the New Vacuum Solar Telescope\footnote{\url{http://fso.ynao.ac.cn}} (NVST; \citealp{liu14,yan20}) and the Solar Dynamics Observatory\footnote{\url{https://sdo.gsfc.nasa.gov}} (SDO; \citealp{pes12}). 
The NVST is a vacuum solar telescope boasting a 985 mm clear aperture, situated at Fuxian Lake in Yunnan Province, China. It could provide high-resolution imaging observations of the photosphere (TiO 7058 $\rm\AA$) to investigate fine structures. The TiO images have a pixel size of 0.\arcsec052 and a cadence of 30 s. The field of view (FOV) for TiO images is 120\arcsec $\times$ 100\arcsec. These data undergo calibration from Level 0 to Level 1, inclusive of dark current subtraction and flat-field correction. Subsequently, the speckle masking method is employed to reconstruct the calibrated images from Level 1 to Level 1+, following the approach outlined in \citet{xia16}.

The Helioseismic and Magnetic Imager (HMI; \citealp{sch12,sche12}) aboard the Solar Dynamics Observatory (SDO) provides full-disk, multiwavelength, high spatio-temporal resolution observations for this study. By inverting the 6173 $\rm\AA$ Fe I absorption line, the SDO/HMI can supply the line-of-sight (LOS) Dopplergram, LOS magnetic field, continuum intensity, and vector magnetic field \citep{hoe14} in the solar photosphere with a spatial size of 0.\arcsec5 pixel$^{-1}$, with the cadences of the three former channels being about 45 s and that of the latter one being about 12 minutes. Doppler velocities are calibrated by removing SDO motion and solar rotation \citep{wel13}. Doppler velocity, LOS magnetic field, and vector magnetic field are employed to reveal the physical information in the photosphere.

\section{Results}
\label{sect:results}
An active region, NOAA 12720, featuring a leading positive sunspot followed by a trailing negative sunspot, gradually appeared in the northern hemisphere from August 24 to 25, 2018. The presence of a leading positive sunspot followed by a negative sunspot challenges the typical hemispheric rule observed for active regions in the northern hemisphere during solar cycle 24 \citep{hal19}. Consequently, active region NOAA 12720 is classified as an anti-Hale active region. For a more detailed evolution of the active region, refer to the study of \cite{wan21}.
 \begin{figure}
   \centering
   \includegraphics[width=\textwidth, angle=0]{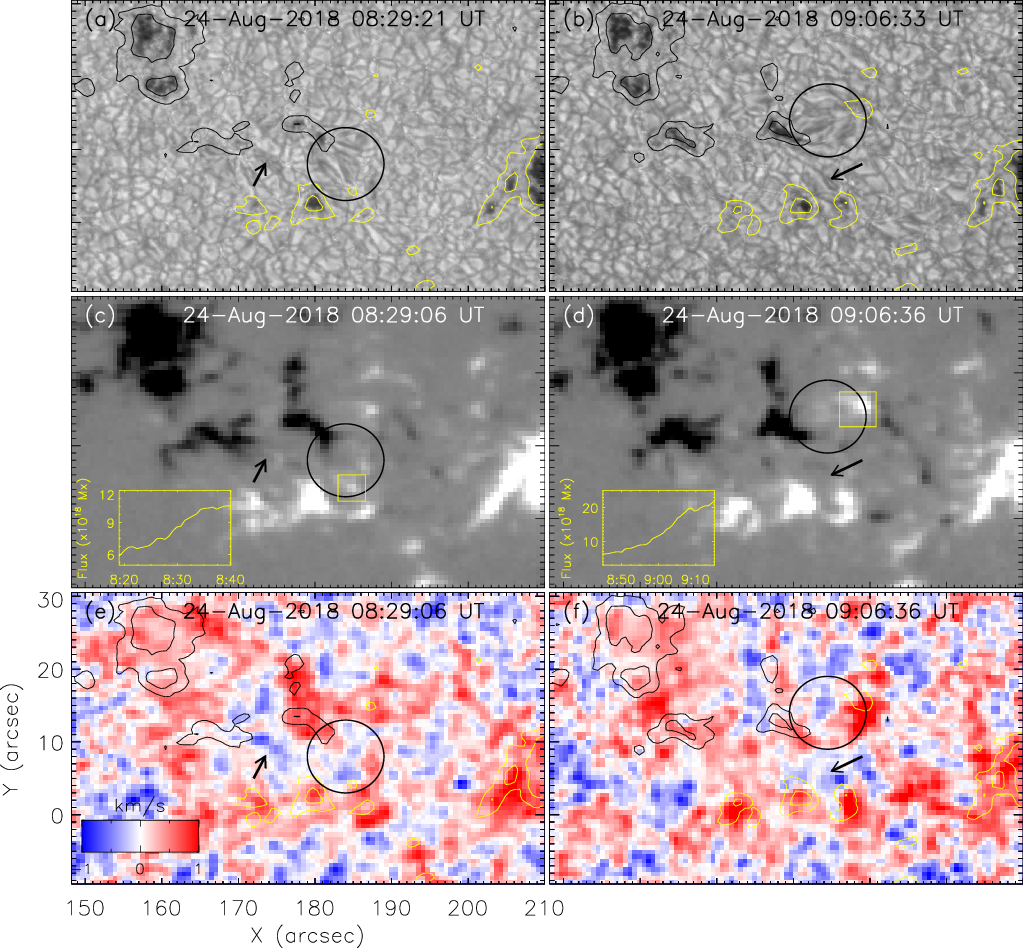}
   \caption{The elongated granulations and stretched dark lances. (a)-(b) TiO images observed by NVST. (c)-(d) Corresponding magnetograms derived from SDO/HMI. White (black) indicates the magnetic field with positive (negative) polarity, saturating at $\pm$ 500 G. (e)-(f) Corresponding Dopplergrams derived from SDO/HMI. Positive (red) and negative (blue) velocity values represent downflows and upflows, respectively. Yellow and black contours delineate the magnetic field with positive and negative polarities at levels of $\pm$ 250 G and $\pm$ 750 G, respectively. Black circles and arrows highlight some elongated granules and stretched dark lanes.}\label{fig1}
   \end{figure}
 
\begin{figure}
   \centering
   \includegraphics[width=\textwidth, angle=0]{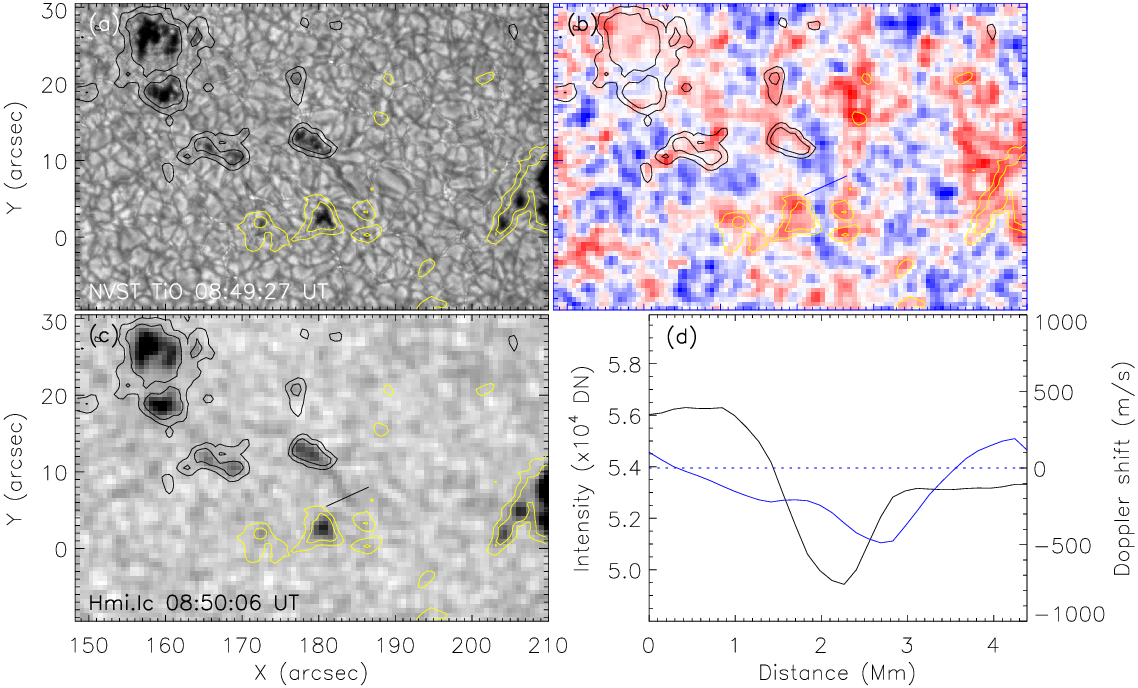}
   \caption{The elongated dark lane observed around 08:49 UT. (a) TiO image. (b) Corresponding Dopplergram. (c) Corresponding continuum intensity map. The yellow and black contours in panels (a)-(c) indicate magnetic fields with positive and negative polarities at levels of $\pm$250 G and $\pm$750 G, respectively. (d) Variations in continuum intensity and Doppler shift along the path marked by the solid line in panels (b) and (c). The black line represents the continuum intensity, while the blue line denotes the Doppler shift.}\label{fig11}
   \end{figure}

During the emergence of the active region, numerous elongated granules and stretched dark lanes were concurrently observed through the TiO passband by NVST (refer to the animation1), occurring between the two primary polarities. Subsequently, some dark pores gradually became apparent. Fig. \ref{fig1} illustrates several elongated granulations and stretched dark lanes in the EFR. The first row displays TiO images observed by NVST, while the second and third rows depict the corresponding magnetograms and Dopplergrams from SDO/HMI, respectively. 
In panels (a) \& (b), black circles and arrows draw attention to particular elongated granulations and stretched dark lanes. Corresponding circles and arrows are also depicted in panels (c) \& (d) and (e) \& (f). Based on the TiO images, the elongated granulations and stretched dark lanes alternately occupied positions, with lengths ranging from 2.7 Mm to 4.2 Mm. Notably, these elongated granulations and stretched dark lanes emerged in the midst of the newly emerging positive and negative polarities and exhibited connections with them (see panels (c) \& (d)).
Additionally, we calculated the positive magnetic fluxes associated with one footpoint of these elongated granulations and stretched dark lanes within the yellow boxes of panels (c) \& (d). The yellow curves in the bottom left of panels (c) \& (d) depict the temporal variations of positive magnetic flux in each yellow box. Notably, the positive magnetic flux exhibited a rapid increase in these regions, indicating an association between the elongated granulations and stretched dark lanes with the newly emerging flux. The magnitude of the newly emerging magnetic flux could reach up to the order of 10$^{19}$ Mx, a strength substantial enough to influence the morphology of the granules in the photosphere \citep{che07}.
Hence, we deduce that the elongated granulations and stretched dark lanes were the consequence of the interaction between the newly emerging flux and ambient material on the solar surface. By examining the corresponding Dopplergrams in panel (e) \& (f), it is evident that the elongated granulations or stretched dark lanes are associated with blueshift. This is particularly clear in the case of the stretched dark lanes, as marked by the black arrows in panels (e) and (f), and as shown in Animation 2. 
Fig. \ref{fig11} presents a case to illustrate this association. Panel (a) shows the TiO image at 08:49 UT, while panels (b) and (c) show the corresponding LOS Dopplergram and Continuum intensity map, respectively. Panel (d) plots the variations of continuum intensity and LOS Doppler shift along the solid lines in panels (b) and (c). We can see that the stretched dark lane is dominated by blueshift.
This observation leads us to speculate that the stretched dark lanes may correspond to the rising flux tube.
However, by carefully examining Animation 2, we could also find that not all the stretched dark lanes are dominated by blue shifts; some are even dominated by red shifts. There may be two reasons for this. First, downflowing plasma along the rising flux tube can lead to red shift signatures when the downflows are faster than the rising flux tube. Second, the low accuracy of SDO/HMI Dopplergrams would introduce some uncertainties. Overall, we consider that these stretched dark lanes are likely associated with the rising flux tubes.

 \begin{figure}
   \centering
   \includegraphics[width=\textwidth, angle=0]{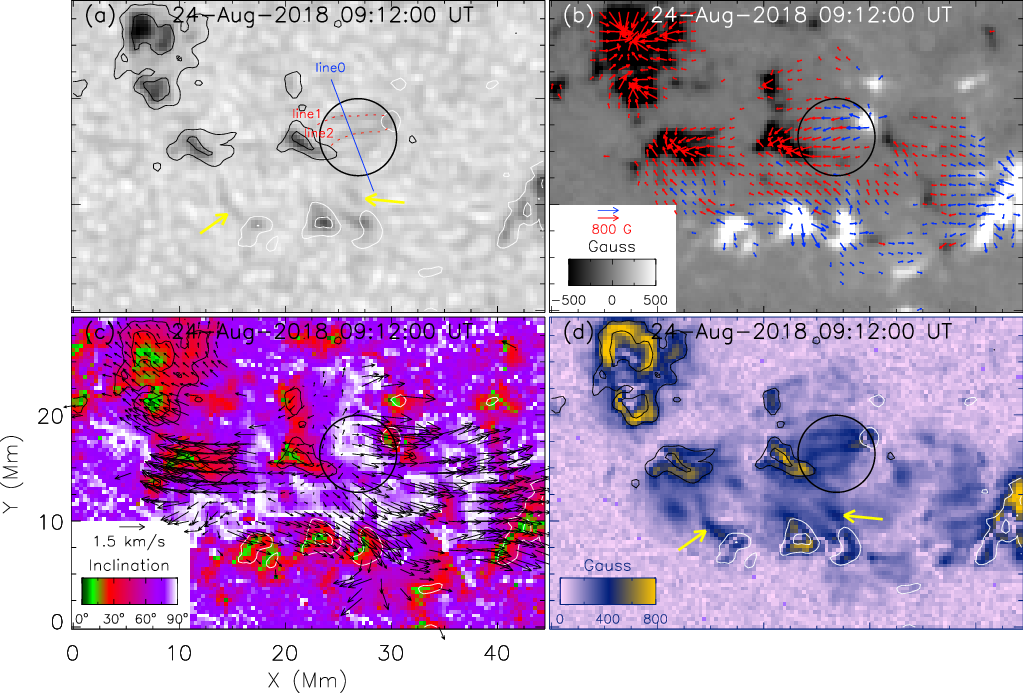}
   \caption{The magnetic properties of elongated granules and stretched dark lanes. (a) Continuum intensity map from SDO/HMI. (b) Vector magnetogram observed by SDO/HMI. Blue and red arrows represent the transverse magnetic field for pixels where the field strength is greater than 200 G with positive and negative polarities, respectively. The positive (negative) vertical field is indicated by the white (black) background. (c) Inclination angle superimposed with transverse velocity field vectors derived by DAVE4VM. (d) Strength of the transverse magnetic field from SDO/HMI. The white and black contours in panels (b), (c) \& (d) delineate the magnetic field with positive and negative polarities at levels of $\pm$ 250 G and $\pm$ 750 G, respectively. The black circle in each panel encompasses the region of elongated granules and stretched dark lanes at 09:06:33 UT on August 24, referring to panel (b) of Fig.\ref{fig1}.}\label{fig2}
   \end{figure}

 \begin{figure}
   \centering
   \includegraphics[width=\textwidth, angle=0]{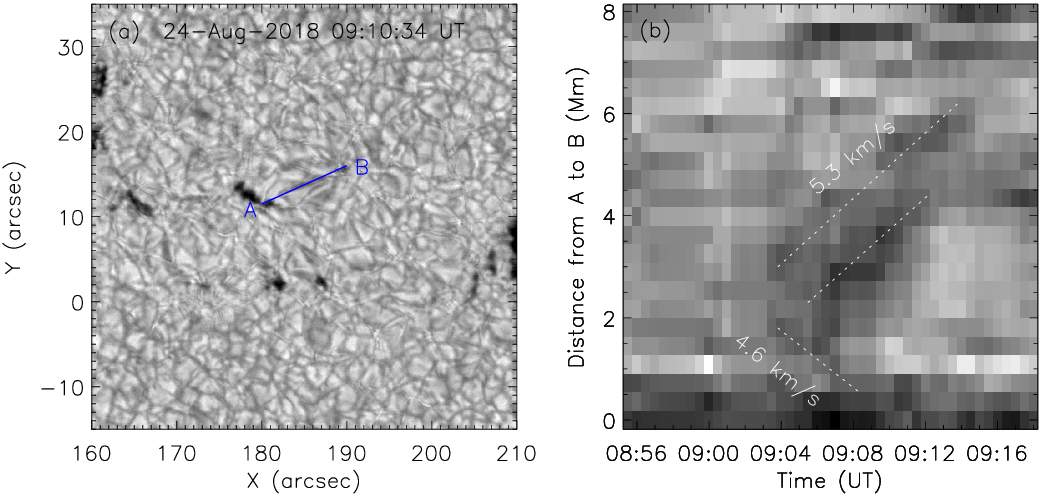}
   \caption{Separation motion around 09:10 UT. (a) TiO image observed by NVST at 09:10 UT. (b) Time-distance diagram along path AB in panel (a), reconstructed from a series of TiO images.}\label{fig31}
   \end{figure}

 Fig. \ref{fig2} displays the photospheric observations around 09:12 UT on August 24, corresponding to the elongated granulations and stretched dark lanes shown in Fig. \ref{fig1} (b). Panel (a) presents the continuum image observation from SDO/HMI data. Some elongated granulations and stretched dark lanes are identified within the black circle, positioned between two distinct polarities (white and black contours) and connected to them.
 Additionally, there are stretched dark lanes occurring in other locations, as indicated by the yellow arrows. These stretched dark lanes also establish connections with two opposing polarities (positive and negative polarities). Panel (b) displays the vector magnetogram from SDO/HMI. The direction of the transverse field is indicated by the arrow's direction, and the length of the arrow represents the strength of the transverse field. The transverse magnetic fields marked by the black circle roughly align with the elongated granulations and stretched dark lanes. 
 Panel (c) illustrates the contribution of the inclination angle superimposed with the transverse velocity field vectors derived by the Differential Affine Velocity Estimator for Vector Magnetograms (DAVE4VM) method \citep{sch08}. The inclination angles are calculated using the formulas: $\theta = \arctan\left(\left|\frac{B_t}{B_r}\right|\right)$, where $B_t$ and $B_r$ represent the transverse and radial magnetic fields. The centers of the elongated granules and stretched dark lanes exhibit large inclination angles, while the two ends of the elongated granules and stretched dark lanes have relatively small inclination angles. The mean transverse field and the mean inclination angle in the black circle are estimated at 300 G and 77$\degr$, respectively. On the other hand, diverging flows can be identified in the region marked by the black circle (observe the black arrows in panel (c)). The flows are directed towards the region where magnetic flux is concentrated, indicating that the two polarities are moving away from each other. The positive polarity moves towards the solar west, while the negative polarity moves in the opposite direction.
 This feature is also consistent with the scenario of flux emergence \citep{che07}. Indeed, as a flux tube ascends into the upper atmosphere from the subsurface, the two footpoints of the lifting flux tube would separate from each other, causing the separation of two polarities in the photosphere. In order to estimate the separation speed, we created a time-distance diagram along the direction of the motion (see Fig. \ref{fig31}). By tracking the trajectory of the dark structure, we determined that the speed towards the solar east is approximately 4.6 km/s, while the speed towards the solar west is about 5.3 km/s. Therefore, the separation speed is approximately 10 km/s.
 
\begin{figure}
   \centering
   \includegraphics[width=\textwidth, angle=0]{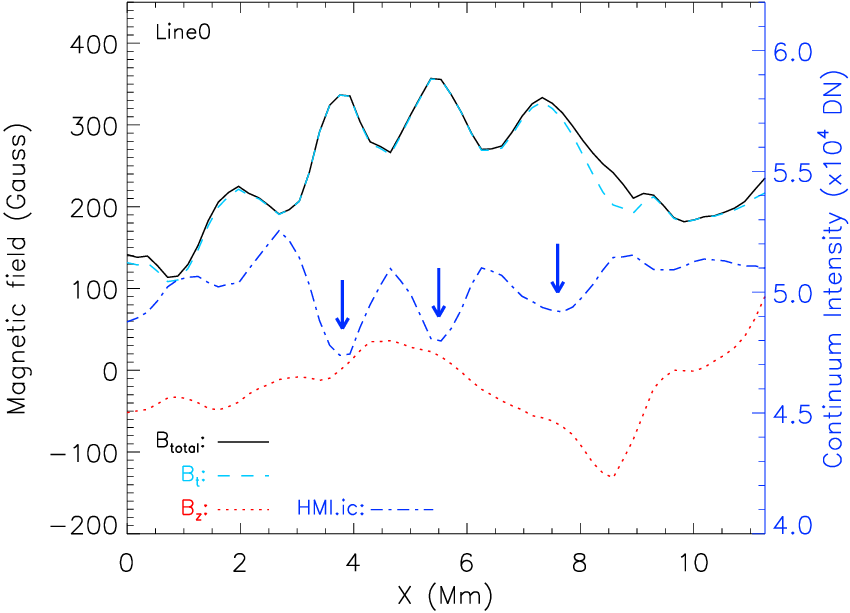}
   \caption{Variations of different parameters along the blue line (line0) in Fig. \ref{fig2}. The starting point of the X-axis is the northeastern end of line0. The variation of the total magnetic field is depicted by a black solid curve, while the continuum intensity variation is represented by the blue dotted-dashed curve. The variations of the transverse and vertical magnetic fields are shown by pink dashed and red dotted curves, respectively.}\label{fig3}
   \end{figure}

Panel (d) of Fig.\ref{fig3} illustrates the strength of the transverse magnetic field calculated by the equation $B_t=\sqrt{B_x^2+B_y^2}$, where $B_x$ and $B_y$ represent the two perpendicular components of the transverse magnetic fields. The intriguing observation is that the stretched dark lanes correspond to regions with a strong transverse magnetic field, as indicated by the black circle and yellow arrows in panels (a) \& (d). The morphologies of the stretched dark lane and the strong transverse magnetic field are quite similar, suggesting a close association between stretched dark lanes and flux tubes with a strong transverse magnetic field.
   
Fig. \ref{fig3} illustrates the variations of different parameters across the elongated granules and stretched dark lanes along the blue line (line0) in Fig. \ref{fig2} (a). The black solid curve represents the total magnetic field ($B_t$), while the blue dotted-dashed curve represents the continuum intensity. The pink dashed and red dotted curves represent the two different components of the magnetic field (transverse ($B_t$) and vertical ($B_z$) magnetic field). The transverse magnetic field exhibits a similar and equivalent variation to the total magnetic field, indicating that the magnetic field along line0 is predominantly influenced by the transverse component. 
Furthermore, there is a negative relationship between the continuum intensity and the total/transverse magnetic field. Importantly, three noticeable ``dips'' marked by the blue arrows are identified in the continuum intensity, corresponding to the stretched dark lanes. These stretched dark lanes exhibit a strong total/transverse magnetic field, while the gap between two adjacent stretched dark lanes shows a lower total/transverse magnetic field (see the black solid and pink dashed curves in Fig. \ref{fig3}). In other words, the center of elongated granules have fewer total/transverse magnetic fields compared to stretched dark lanes.
\begin{figure}
   \centering
   \includegraphics[width=\textwidth, angle=0]{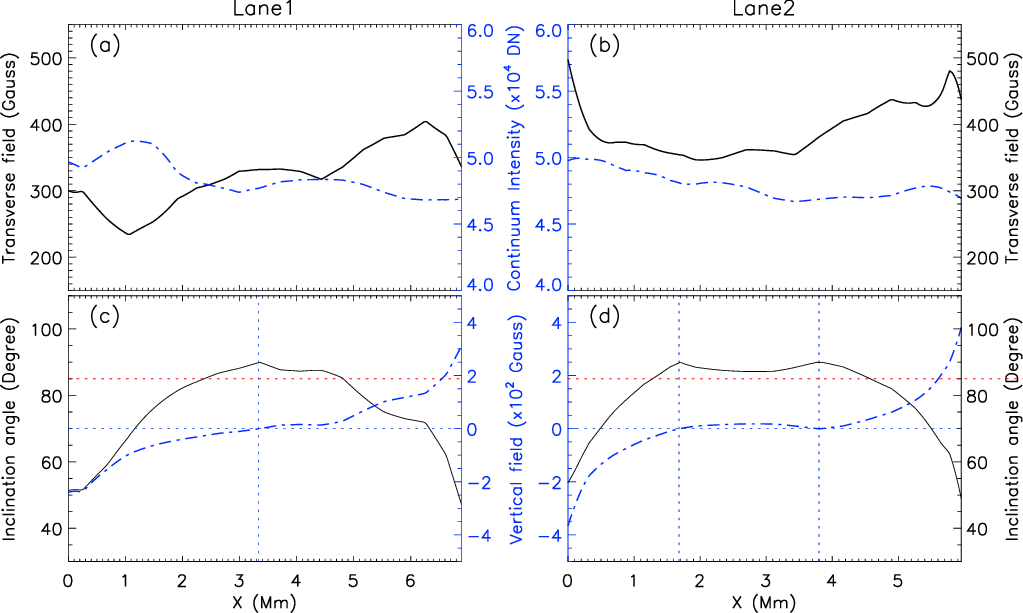}
   \caption{Variation of the transverse magnetic field, continuum intensity, inclination, and vertical magnetic field along two different stretched dark lanes marked by a red dotted line in Fig. \ref{fig2}(a). (a) \& (b) Transverse magnetic field (dark solid curve) and continuum intensity (blue dotted-dashed curve) for two different lanes. (c) \& (d) Inclination angle (black solid curve) and vertical magnetic field (blue dotted-dashed curve) for two different lanes. The blue horizontal dotted line denotes the zero level of the vertical magnetic field in panels (c) \& (d), while the blue vertical dotted lines mark the zero site of the vertical magnetic field.}\label{fig4}
   \end{figure}

To study the stretched dark lane in detail, we also analyze the magnetic properties along two different stretched dark lanes (named Lane1 and Lane2) marked by two red dotted lines (line1 and line2 in Fig. \ref{fig2} (a)). The left column of Fig. \ref{fig4} represents the variations of different parameters for Lane1, while the right column is for Lane2. In each panel, the beginning of the X-axis corresponds to the left end of a line1 or line2 in Fig. \ref{fig2}. Panels (a) \& (b) show the variations of the transverse magnetic field and continuum intensity along the two stretched dark lanes, respectively. The black solid curve represents the transverse magnetic field, while the blue dotted-dashed curve represents the continuum intensity. For Lane1, the transverse magnetic field ranges from 230 G to 400 G, exhibiting a negative relationship with the continuum intensity. For Lane2, the transverse magnetic field is about 350 G to 490 G, which is higher than that of Lane1. The transverse magnetic field in the middle of this lane could be higher than 350 G.
However, an obvious negative relationship between the transverse magnetic field and continuum intensity in this lane cannot be identified. Panels (c) \& (d) display the variations of the inclination angle and vertical magnetic field, respectively. The black solid curve represents the inclination angle, while the blue dotted-dashed curve represents the vertical magnetic field. For Lane 1, the inclination angle exhibits an ``arcade'' pattern, increasing initially and then decreasing. Additionally, the vertical magnetic field in the left part is negative, while in the right part, it is positive. Moreover, according to Fig. \ref{fig2} (b), we can infer that the directions of the transverse magnetic field in this place are uniform. Therefore, we can deduce that the left part of this lane has the opposite polarity of inclination angles as the right part. The magnetic fields are almost horizontal in the middle part, with inclination angles exceeding 85$\degr$ as marked by the red dotted line. For Lane 2 (see panel (d)), the inclination angle also follows an ``arcade'' pattern, similar to Lane 1. The magnetic fields are also approximately horizontal in the middle part marked by the red dotted line. The vertical magnetic field shows the same pattern as Lane 1, with the left part being negative and the right part being positive. These features indicate that an arcade-like shaped flux tube lies on the stretched dark lane.

\section{Summary and discussion}\label{sec:conclusion}
In this paper, we focus on the elongated granules and stretched dark lane associated with the emergence in a purely naked emerging anti-Hale active region NOAA 12720. The main results are as follows

(1) Many elongated granules and stretched dark lanes inhabited alternately each other with the lengths ranging from 2.7 Mm to 4.2 Mm in the EFR, which were associated with the newly emerging flux. The blue shift in the middle of the elongated granules and stretched dark lanes are found, and the stretched dark lanes have a close correlation with a strong blue shift in the EFR.

(2) Strong transverse magnetic field in the region of elongated granules and stretched dark lanes is found, which also aligns with the direction of elongated granules and stretched dark lanes. On the other hand, separative motions are observed at the two ends of the elongated granules and stretched dark lanes, with the speed of these motions estimated to be about 10 km/s.

(3) The stretched dark lanes instead of elongated granules show a close relationship with the strong magnetic field, which manifests that stretched dark lanes may be the signatures of the lifting magnetic flux tube. The transverse magnetic field of stretched dark lanes could be up to around 300 to 400 G. The ``arcade'' pattern of the magnetic field in the stretched dark lanes is found.

\begin{figure}
   \centering
   \includegraphics[width=\textwidth, angle=0]{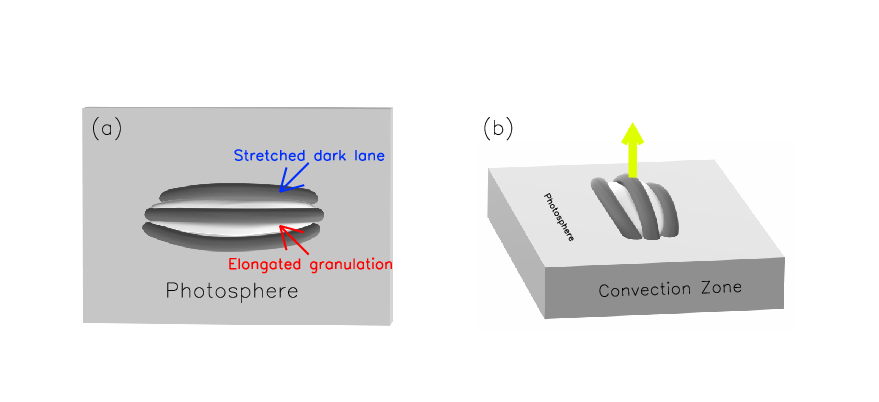}
   \caption{Cartoon showing the elongated granulations and stretched dark lanes seen from a top view (panel (a)) and side view (panel (b)).}\label{fig5}
   \end{figure}

Many glamorous phenomena often occur in the EFR, such as elongated granules, Ellerman bombs, arch filament systems, and so on, which are regarded as the result of the interaction between the emerging magnetic field and ambient magnetic field or plasma \citep{che07,shen22}. By using high-resolution TiO images observed by the NVST, we also find a lot of elongated granules and stretched dark lanes inhabiting the EFR. They are arranged alternately with a strong transverse field and high inclination angle, which are associated with newly emerging magnetic flux. Unlike previous studies of emerging magnetic flux in quiet Sun \citep{ort14,kon20}, we mainly pay more attention to the stretched dark lanes instead of elongated granulations. The interesting question is what is the nature of these elongated granules and stretched dark lanes? To our knowledge, the subsurface magnetic fields lift according to buoyant instability, which is in the form of a magnetic flux tube with low density. When these flux tubes with low density go through the photosphere, they push the photospheric material away and continue to ascend. In the emerging process of flux tubes, the magnetic pressure is comparable with convective motion, even predominant at the bottom of the photosphere. This is consistent with the results by \cite{stein06}, which show that near the solar surface, the density inside the flux tube is less than its surroundings and the field reaches pressure balance with the surrounding plasma. Otherwise, the strong magnetic field would influence or constrain the convection of plasma \citep{fan09}. \cite{che07} demonstrated that the emergence process can modify the local granulation pattern depending on the properties of the flux tube (e.g., magnetic strength, magnetic twist). This may be the reason that the emerging magnetic flux can restructure the shape of granulation. Therefore, the elongated granules would be thought of as the gap of the emerging magnetic flux tube with substantial material, whereas the dark lanes would be thought of as a horizontal emerging flux tube with thinner material (see Fig.\ref{fig5} (a) \& (b)). From Fig.2 \& 3 of \cite{cen17}, we also can find that the strong blue shift and horizontal magnetic field mainly inhibited in the stretched dark lane instead of elongated granule. This finding further supports our conjecture. More recently, the high-resolution observations from the Goode Solar Telescope (GST) at Big Bear Solar Observatory, \cite{wan20b} found both types of flux emergence (magnetic loop emergence and flux sheet emergence \citep{mor18}) are associated with darkening of granular boundaries. These observational evidence prove that magnetic flux emergence is closely corrected to stretched dark lanes. 

The footpoints of rising flux tubes could appear as bright points due to the increasing heating from the falling plasma along the flux tube \citep{kel04,ste05,liu18}. However, these types of transient darkenings often correspond to the crest of emerging flux tubes \citep{che07, cheu08}, which is consistent with our observations. Another question is why the crests of these rising flux tubes appear as stretched dark lanes. Rising flux tubes can affect the thermal structure of the surrounding plasma. As these tubes rise, they can influence the local temperature and density, which in turn affects the optical depth \citep{spr06}. In regions where the crests of rising flux tubes are present, the horizontal magnetic field can suppress convection, leading to a decrease in temperature \citep{fan09,che07}. On the other hand, cooling of the plasma inside the crest of a rising flux tube, due to thermal radiation, would also lead to an increase in density \citep{park79,cho92}. This can result in a higher optical depth, causing the region to appear darker. Furthermore, more observational information is needed to understand the nature of elongated granules and stretched dark lanes in the future, especially regarding high-resolution magnetic fields.

\begin{acknowledgements}
We appreciate the referee’s careful reading of the manuscript and many constructive comments, which helped greatly in improving the paper.
SDO is a mission of NASA's Living With a Star Program. The authors are indebted to the SDO, and NVST teams for providing the data. This work is supported by the National Key R \& D Program of China (2019YFA0405000), the Strategic Priority Research Program of the Chinese Academy of Sciences, Grant No.XDB0560000, the National Science Foundation of China (NSFC) under grant numbers 12003064, 12325303, 11973084, 12203020, 12203097, 12273110, 12003068, the Yunnan Key Laboratory of Solar Physics and Space Science (202205AG070009), the Yunnan Science Foundation of China under number 202301AT070347, 202201AT070194, 202001AU070077, Yunnan Science Foundation for Distinguished Young Scholars No. 202001AV070004, the grant associated with a project of the Group for Innovation of Yunnan province.
\end{acknowledgements}

\label{lastpage}
\end{document}